\documentclass[twocolumn,showpacs,preprintnumbers,amsmath,amssymb]{revtex4}

\usepackage{graphicx}
\usepackage{dcolumn}
\usepackage{bm}
\usepackage{epstopdf}
\usepackage{color}

\begin{document}


\title{Velocity-selective two-photon absorption \\
induced by a diode laser in combination with a train of ultrashort pulses}

\author{Marco~P.~Moreno}
\affiliation{
Departamento de F\'{\i}sica, Universidade Federal de Rond\^onia,
 76900-726 Ji-Paran\'a, RO - Brazil}

\author{Giovana~T.~Nogueira}
\affiliation{
Departamento de F\'{\i}sica, Universidade Federal de Juiz de Fora, \\
36036-330 Juiz de Fora, MG - Brasil
}%

\author{Daniel~Felinto}
\author{Sandra~S. Vianna}%
 \email{vianna@ufpe.br}
\affiliation{%
Departamento de F\'{\i}sica, Universidade Federal de Pernambuco,  \\
50670-901 Recife, PE - Brazil
}%

\date{\today}

\begin{abstract}

The two-photon transition $5S-5P-5D$ in rubidium vapor is investigated by detecting the fluorescence from the $6P_{3/2}$ state when the atomic system is excited by the combined action of a cw diode laser and a train of ultrashort pulses. The cw-laser plays a role as a velocity-selective filter and allows for a spectroscopy over a large spectral range including the $5D_{3/2}$ and $5D_{5/2}$ states. For a counterpropagating beam configuration, the response of each atomic velocity group is well characterized within the Doppler profile, and the excited hyperfine levels are clearly resolved. The contribution of the optical pumping to the direct two-photon process is also revealed. The results are well described in a frequency domain picture by considering the interaction of each velocity group with the cw laser and the modes of the frequency comb.
\end{abstract}

\pacs{Valid PACS appear here}
\maketitle

\section{\label{sec:1}Introduction}

In the last decades, mode-locked femtosecond (fs) lasers have been established as an important tool for atomic and molecular spectroscopy, with applications in the fields of biology, chemistry and physics \cite{Udem2002, Ye2005, Peer2007, Nishiyama2015}. With a spectrum consisting of evenly spaced narrow lines, the phase-controlled wide-bandwidth optical frequency comb provides a precise and direct link between microwave and optical frequencies \cite{Udem2002}, and can cover from extreme UV \cite{Altmann2016} to the mid-IR \cite{Vainio2016}. In these applications, a great variety of  new sets of techniques have been developed, usually based on the use of the frequency comb either as a rule to measure the frequency of a cw laser, which interacts with an atomic transition \cite{Udem1999, Chui2005}, or as a single direct probe of an atomic or molecular transition \cite{Marian2004}. Specifically, for a direct frequency-comb spectroscopy, the two-photon transition is commonly employed, by exploring the fact that the resonance condition can be simultaneously satisfied by many pairs of comb lines \cite{Stalnaker2010, Barmes2013, Hipke2014}. An interesting example along these lines is the recently demonstration \cite{Jayich2016} that the entire spectrum of an optical frequency comb can cool and trap atoms when used to drive a narrow two-photon transition.

In this work, we focus on a different scheme, based on the introduction of a second, cw laser that works in combination with the femtosecond laser, so that both interact with the atomic system. In this scheme, the narrowband laser assumes the role of a velocity-selective filter, opening new directions of investigation, in special, for atomic systems with considerable Doppler broadening, where coherent accumulation processes are present \cite{Felinto2003}. Previous investigations of one-photon and two-photon transitions with this scheme have already been performed \cite{Aumiler2005, Moreno2012}. In the case of one-photon transition, the cw laser probe the action of the femtosecond laser over the various velocity groups, resulting in the frequency comb printed on the Doppler profile, due to the velocity distribution of the excited state population, and in the velocity-selective population transfer between the atomic ground-state hyperfine levels \cite{Aumiler2009}. When a fs laser with 1 GHz frequency separation of the optical modes is applied to investigate an atomic vapor at room temperature, only one mode can fit within the Doppler profile, allowing  us to distinguish the different hyperfine levels \cite{Polo2011}. The situation is different for the two-photon transition, when the cw laser is responsible for driving one of the steps of the excitation  process. In particular, for the $5S \rightarrow 5P \rightarrow 5D$ two-photon transition in rubidium vapor, previous studies using contra- and co-propagation beam configurations revealed well resolved hyperfine levels of the $5D$ state (contra-propagating beams) \cite{Moreno2012} and the frequency comb, that drives the upper transition,  printed in the excitation spectra of the blue fluorescence (co-propagating beams) \cite{Lira2015}. A similar scheme \cite{Moon2011} using the double-resonance optical pumping spectroscopy have also been applied to measure the 5P-4D transition in Rb.

Here, we present an extension of our previous study on the combined action of a train of ultrashort pulses and a cw diode laser over the $5S \rightarrow 5P \rightarrow 5D$ two-photon transition in rubidium vapor \cite{Moreno2012}. Using a contra-propagating beams configuration, we explore the selectivity in the velocity and analyze the response of different groups of atoms within the Doppler profile. By setting the repetition rate of the fs laser and varying the diode frequency we probe the different groups of atoms that can interact with a fixed mode of the frequency comb. On the other hand, we can also probe a specific group of atoms each time by selecting a fixed diode frequency and scanning the repetition rate.
In  particular, the former scheme allows us to use the saturated absorption signal of the diode lasers as a frequency guide inside the Doppler profile and to associate the fluorescence signal to each atomic velocity group. In addition, working with a fs laser with 1 GHz high repetition rate, the necessary condition for the accumulation
of population and coherence is easily fulfilled, and a good
description of the results is obtained considering a three level
cascade system interacting with a cw laser and a single mode of the frequency comb.

Moreover, our results also reveal the contribution of the optical pump, induced by the diode laser, to the direct two-photon absorption process due only to the fs laser.  In this case, broad peaks are observed and theoretical calculations including two modes of the frequency comb are performed.

In the following, we introduce our experimental setup in Section \ref{sec:2} together with the central experimental results. In Section \ref{sec:3} we present our model for the experiments of Section \ref{sec:2}, taking  into account power broadening effects. To validate our approximations in the frequency-domain treatment we compare that with the results obtained in the time domain, considering the whole train of ultrashort pulses. We also compare the experimental e calculated responses of different atomic velocity groups and, analyze the influence of the optical pumping to the fluorescence signal. Finally, in Section \ref{sec:4} we present our conclusions.

\section{\label{sec:2}Experiment}

Our experimental setup is schematically illustrated in Fig.~1 together with the relevant energy levels. 
A diode laser, stabilized in temperature and with a linewidth of about $1$~MHz, is used to excite 
the $5S_{1/2} \rightarrow 5P_{3/2}$ transition at $780$~nm. A train of fs pulses generated by a mode-locked Ti:sapphire laser (BR Labs Ltda) can excite both $5S_{1/2} \rightarrow 5P_{3/2}$ and $5P_{3/2} \rightarrow 5D$ transitions. The two beams 
are overlapped, with orthogonal linear polarizations and in a counterpropagating configuration, 
in the center of a sealed Rb vapor cell. The vapor cell is heated to 
$\approx 80~^0$C and contains both $^{85}$Rb and $^{87}$Rb isotopes in their natural abundances. 

\begin{figure}[ht]
  \centering
  \includegraphics[width=7.0cm]{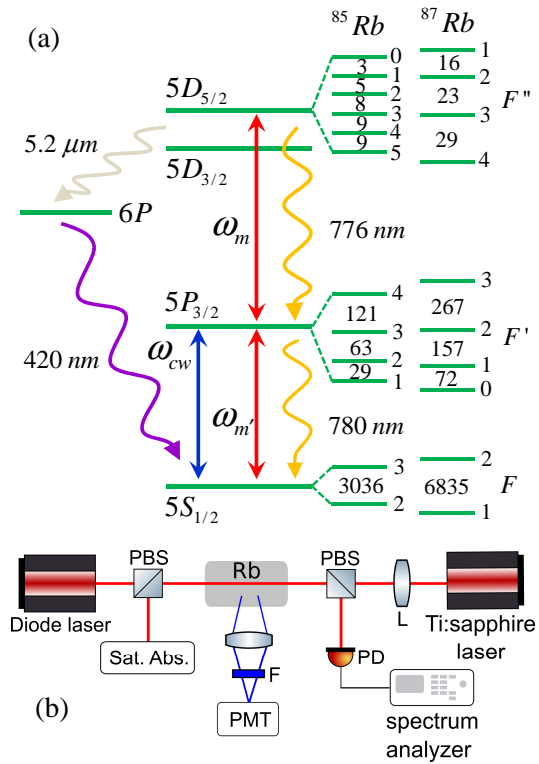}
  \caption {(Color online) (a) Schematic representation of the energy levels of Rb that are relevant for the experiment. The hyperfine splittings are in units of MHz. (b) Experimental setup, where the symbols PMT, PBS, PD and F stand for the photomultiplier tube, polarizer beam splitter, photodiode and filter, respectively.}
  \end{figure} 

The Ti:sapphire laser produces $100$~fs pulses and $300$~mW of average power, such that the power
per mode is $\approx 60$~$\mu$W. The $f_{R}=1$~GHz repetition rate is measured with a photodiode and
phase locked to a signal generator (E8663B-Agilent), with $1$ Hz resolution, while the
carrier-envelope-offset frequency, $f_{0}$, is left free. The diode laser can sweep over
$10$~GHz by tuning its injection current and a saturated absorption setup is used to
calibrate its frequency. A direct detection of the diode-beam transmission after passing through
the cell gives information about the absorption in the $5S \rightarrow 5P$ transition. The diameter of the
two beams at the center of the cell is on the order of $250$~$\mu$m for the fs beam and $1.8$~mm for
the diode laser, leading to a common interaction time of about $800$~ns (corresponding to a linewidth of $\gamma\approx 2\pi\times200$~kHz).

The fluorescence at 420 nm emitted by spontaneous decay from the 6P state to the 5S is collected at $90^0$, 
using a $10$~cm focal lens and a filter to cut the scattered light from the excitation laser beams.
The signal is detected with a photomultiplier tube and recorded on a digital oscilloscope.
Figure 2 shows the fluorescence signal (midle/red curve), for a fixed $f_{R}$, as the 
diode frequency is scanned over the four Doppler-broadened $D_{2}$ lines of the $^{85}$Rb and 
$^{87}$Rb. The spectrum consists of several narrow peaks over a flat background. The 
narrow peaks are due to the two-photon transition excited by both lasers: the diode laser and the 
different modes of the frequency comb; while the background is due only to excitation by
the frequency comb. In the same scan we can observe, simultaneously, peaks associated with the 
excitation of the $5D_{3/2}$ and $5D_{5/2}$ states. We also see two peaks, separated by one 
$f_{R}$ in optical frequency of the diode laser, that correspond to the same 
transition excited by two neighboring modes of the frequency comb. We also present a measurement of the direct transmission of the diode beam after passing through the warm cell (lower/black curve), indicating the strong absorption in the center of the Doppler lines. 
The saturated absorption curve (upper/blue curve) is used to calibrate the diode frequency. All the three curves are detected simultaneously.

We can also detect the fluorescence signal as a function of the repetition rate for a fixed frequency of the diode laser (free-running) as shown in Fig. 3.  The blue curve represents the fluorescence signal without the presence of the diode laser and, therefore, due solely to the fs laser (corresponding to the background signal in figure 2). In this case, as the two absorbed photons are co-propagated we have a Doppler-broadened line. We see that the fluorescence signal is always present,  which indicates that it is always possible to find a mode in resonance with the upper transition at the same time that another mode is at resonance with at least one Doppler line for some atomic-velocity group . Besides that, the intensity of the fluorescence varies slowly as the repetition rate change, with an enhancement not bigger than a factor of two near the double-resonance condition, when the energy difference between two consecutive transitions is a multiple of $f_{R}$ for some atomic-velocity group \cite{Ban2013}.  

The red curve in Fig. 3 represents the blue fluorescence signal now in the presence of the diode laser with a fixed frequency in one of the Doppler lines. In this case, as $f_{R}$ is varied, a peak is observed when the frequency of one mode of the frequency comb plus the fixed diode frequency is equal to the frequency of the two-photon transition. This occurs at intervals of $\approx 2\pi f_R^2/\omega_{5P-5D} \approx 2.6$ kHz \cite{Ban2013}, corresponding to a change in the one-photon optical frequency  of $\approx$ 1 GHz, where $\omega_{5P-5D}$ is the frequency of the $5P_{3/2} \rightarrow 5D_{5/2}$ transition.

\begin{figure}[ht]
  \centering
  \includegraphics[width=7.5cm]{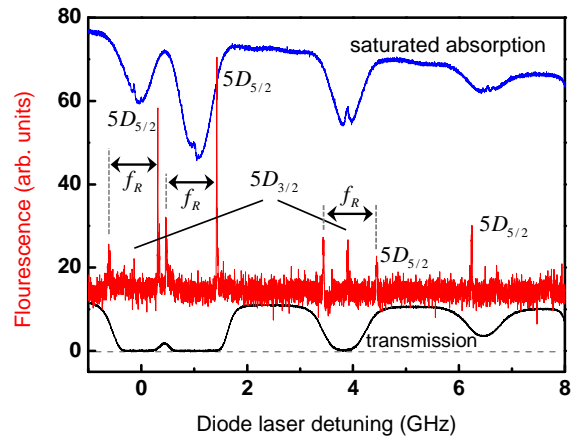}
  \caption {(Color online) Fluorescence from the $6P_{3/2} \rightarrow 5S_{1/2}$ decay as a function of the diode laser frequency for the four $D_{2}$ Doppler lines (middle/red curve) with $f_{R}$ = 1.004\:411\:920 GHz. The saturated absorption signal (upper/blue curve) and the diode transmission after the Rb cell (lower/black curve) are detected simultaneously with the fluorescence signal. Zero detuning is chosen at the $^{87}$Rb, $5S, F_{g} = 2 \rightarrow 5P_{3/2}, F' = 3$ transition.}
  \end{figure} 

\begin{figure}[ht]
  \centering
  \includegraphics[width=6.5cm]{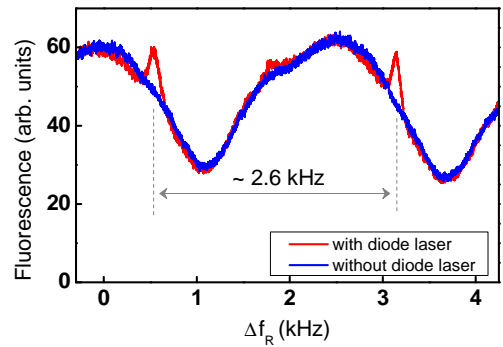}
  \caption {(Color online) The $6P_{3/2} \rightarrow 5S_{1/2}$ fluorescence as a function of fs laser repetition rate variation ($\triangle f_{R}$), with and without the presence of the diode beam (tuned in the $^{87}$Rb, $F = 2 \rightarrow F'$).}
  \end{figure} 

The fluorescence for the Doppler lines $F=2$ of $^{87}$Rb and $F=3$ of  $^{85}$Rb, as a function of the diode frequency, 
is shown in more details in Fig. 4. In figure 4(a) we compare the fluorescence signal, in the same frequency interval of the diode laser, for two values of the repetition rate, separated by 1800 Hz and with $f_{R} = 1.004\:411\:950$ GHz for the upper (red) curve. The middle curve is the saturated absorption. It is clear that the intensity, frequency position and the number of peaks depend of the repetition rate. This is because the frequency of the modes responsible for the upper transition changes with the repetition rate, so the peaks appear only when the diode frequency plus one mode frequency is equal to the difference of frequency between the 5S and 5D states. A zoom of each one of the four peaks labeled on the two curves in Fig. 4(a) is displayed in figures 4(b) and (c) for excitation from the hyperfine ground state $F=2$ of $^{87}$Rb, and in figures 4(d) and (e) from the ground state $F=3$ of $^{85}$Rb. As we can see, each peak in Fig. 4(a) [also in Fig. 2] consists of a group of peaks, which are the result of excitation from a given ground-state hyperfine level to all the hyperfine levels ($F''$) of the excited-state.

\begin{figure}[ht]
  \centering
  \includegraphics[width=7.5cm]{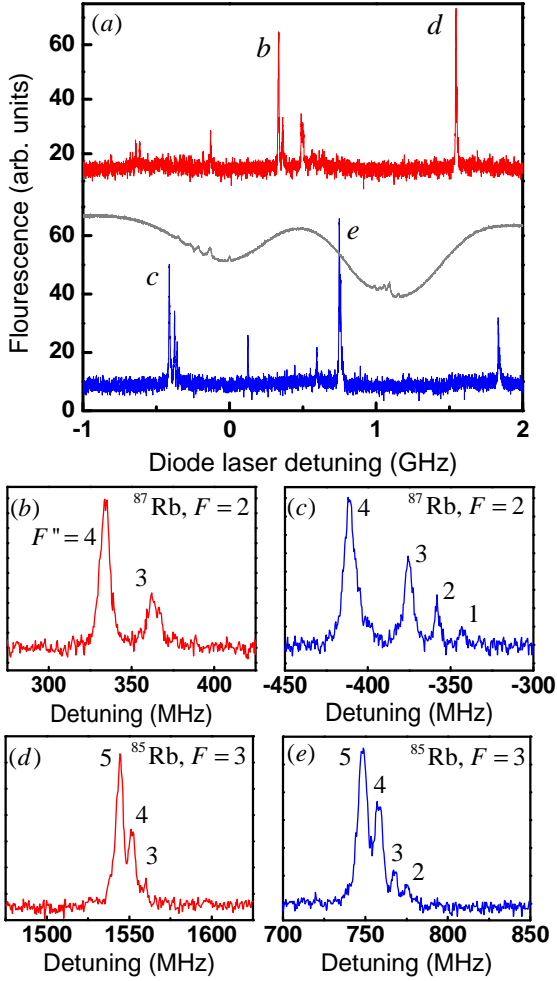}
  \caption {(Color online) (a) Fluorescence at 420 nm as a function of the diode frequency for two values of the repetition rate separated by 1800 Hz. The middle curve is the saturated absorption signal. Zooms of the peaks labeled on the two curves in (a) for excitation to the $5D_{5/2}$ level from the hyperfine ground states: (b) and (c) $F = 2$ of $^{87}$Rb, and (d) and (e) $F = 3$ of $^{85}$Rb. The numbers at the peaks denote the hyperfine final state of the transitions ($F''$).}
  \end{figure} 
  
  \section{\label{sec:3}Theory}

In this section, we present our theoretical model starting with the Bloch equations in Sec. \ref{Bloch} where we introduce the main approximations to obtain an analytical expression that will be used to describe our experimental data. The modeling and the velocity selective process is discussed in Sec. \ref{modeling}, along with a comparison with results obtained by numerical calculation in the time domain, considering the whole train of ultrashort pulses. In the Sec. \ref{OP} we investigate the influence of the optical pumping in the two-photon absorption process.

\subsection{\label{Bloch}The Bloch equations and preliminary considerations}

In order to explain the experimental results presented in the previous section we use a simple model consisting of independent three-level cascade systems interacting with the two fields as schematized in Fig. 5. We denote one of the hyperfine levels 
of the ground ($5S_{1/2},F$), intermediate ($5P_{3/2},F'$), and 
final ($5D,F''$) states as $\left|1\right\rangle$, $\left|2\right\rangle$, and $\left|3\right\rangle$, respectively. The cw diode laser field drives the lower transition $\left|1\right\rangle \rightarrow \left|2\right\rangle$ ($5S$ $\rightarrow$ $5P_{3/2}$), with the fs laser driving the upper transition $\left|2\right\rangle \rightarrow \left|3\right\rangle$ ($5P_{3/2}$ $\rightarrow$ $5D$). As we are interested in the combined action of the diode and the fs lasers, we neglect the background, by assuming that the fs field does not excite the transition 
$\left|1\right\rangle \rightarrow \left|2\right\rangle$.

\begin{figure}[ht]
  \centering
  \includegraphics[width=5.0cm]{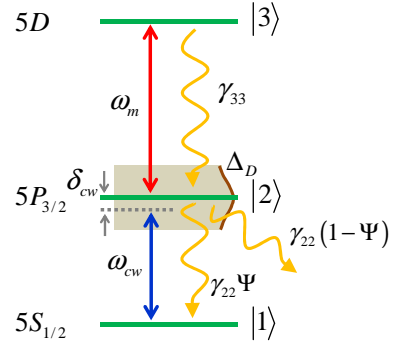}
  \caption {(Color online) Schematic representation of the cascade three-level system used to model the experimental results.}
  \end{figure}

The Hamiltonian of the system is given by $\hat{H} = \hat{H}_0 + \hat{H}_{int}$, where $\hat{H}_0 = \hbar \omega_{21} |2\rangle \langle 2| + \hbar \omega_{32} |3\rangle \langle 3|$
represents the Hamiltonian for the free atom, with the transition frequencies $\omega_{ij} = (E_i-E_j)/\hbar$ and $E_i$ the energy of the $i$-th level. The coupling $\hat{H}_{int}$ describing the interaction between each atom and the two lasers is
\begin{equation}
\label{hamiltonian}
\hat{H}_{int} = -\mu_{12} E_{cw}(t)|1\rangle \langle 2|-\mu_{23} E_{fs}(t)|2\rangle \langle 3| + h.c.,
\end{equation}

\noindent where $\mu_{ij}$ is the dipole moment of the $\left|i\right\rangle\rightarrow \left|j\right\rangle$ transition, and $E_{cw}(t)$ and $E_{fs}(t)$ 
 are the cw and fs electric fields with frequencies $\omega_{cw,fs}$, described by the equations
\begin{equation}
\label{cw-field}
E_{cw}(t) = {\cal E}_{cw}e^{i\omega_{cw} t},
\end{equation}

\begin{equation}
\label{fs-field}
E_{fs}(t) = \sum_{n=0}^{N-1}{\cal E}_{fs}(t-nT_R)e^{i\omega_{fs} t}.
\end{equation}

\noindent Here, ${\cal E}_{cw}$ is the cw field amplitude, ${\cal E}_{fs}(t)$ is the pulse envelope of the fs field, and $T_R=1/f_{R}$ and $N$ are the repetition period and the number of pulses.

The train of ultrashort pulses is described in the frequency domain as a frequency comb and we take into account only the modes that are close to resonance with the 
$\left|2\right\rangle \rightarrow \left|3\right\rangle$ transitions.  
The Rabi frequencies for the diode laser and for each mode of the frequency comb are defined as:

\begin{subequations}
\label{rabi-frequencies}
\begin{align}
\Omega_{cw} &= \frac{\mu_{12}{\cal E}_{cw}}{\hslash},
\\
\Omega_{m} &= \frac{\mu_{23}{\cal E}_{m}}{\hslash},
\end{align}
\end{subequations}

\noindent where ${\cal E}_m$ is the amplitude of the m-th mode of the frequency comb \cite {Moreno2014}. In this notation, the frequency of each mode \textit{m} is given by $\omega_{m}=2\pi(f_{0}+mf_{R})$, where $f_{0}$ is the offset frequency. 

The Bloch equations for a group of atoms with velocity $v$ in the rotating wave approximation are given by:	

\begin{subequations}
\label{bloch}
\begin{align}
\dot{\rho}_{11} &= -\Omega_{cw}\sigma_{12} + c.c. + \Psi\gamma_{22}\rho_{22} - \Gamma(\rho_{11} - \rho^{(0)}_{11}) ,
\\
\dot{\rho}_{22} &= \;\;\:\Omega_{cw}\sigma_{12} + c.c. - \Omega_{m}\sigma_{23} + c.c. \nonumber
\\
                & - (\gamma_{22} + \Gamma)\rho_{22} + \gamma_{33}\rho_{33} , 
\\
\dot{\rho}_{33} &= \;\;\:\Omega_{m}\sigma_{23} + c.c. - (\Psi\gamma_{33} + \Gamma)\rho_{33} - \gamma_{33}\rho_{33} ,
\\
\dot{\sigma}_{12} &=  \left[i\delta_{cw} - \gamma_{12} - \Gamma \right]\sigma_{12} - i\Omega_{m}\sigma_{13} \nonumber
\\
									&+ i\Omega_{cw}(\rho_{22} - \rho_{11}), 
\\
\dot{\sigma}_{23} &=  \left[i\delta_{m} - \gamma_{23} - \Gamma \right]\sigma_{23} - i\Omega_{cw}\sigma_{13} \nonumber
\\
									&   + i\Omega_{m}(\rho_{33} - \rho_{22}), 
\\
\dot{\sigma}_{13} &=  \left[i(\delta_{cw}  + \delta_{m})  - \gamma_{13} - \Gamma\right]\sigma_{13} \nonumber
\\
									&+ i\Omega_{cw}\sigma_{23} - i\Omega_{m}\sigma_{12}, 
\end{align}
\end{subequations}

\noindent where $\rho_{\textit{kl}}$  represents the element $\textit{kl}$ of the atomic density matrix and $\gamma_{\textit{kl}}$ represents its relaxation time. The finite interaction time due to the escape of atoms from the interaction region is accounted by the relaxation rate $\Gamma$. This loss of atoms is compensated by the arrival of new atoms in the ground state at the same rate, and $\rho^{0}_{11}$ is the ground state population in thermal equilibrium. We also consider that the $\left|1\right\rangle \rightarrow \left|2\right\rangle$ 
transitions, driven by the diode laser, may be closed ($\Psi=1$) or open ($\Psi=1/2$), depending on which $F'$ is being excited. However, due to the longer lifetimes of the $5D$ states (resulting in weaker optical pumping) and to simplify the calculations, we consider that the 
$\left|2\right\rangle \rightarrow \left|3\right\rangle$ transitions are always closed. 

The coherences are represented in terms of their slowly varying envelops: $\sigma_{12}=\rho_{12}e^{-i\omega_{cw}t}$, $\sigma_{23}=\rho_{23}e^{-i\omega_{m}t}$ and $\sigma_{13}=\rho_{13}e^{-i(\omega_{cw}+\omega_{m})t}$. The two detunings, taking into account the inhomogeneous Doppler broadening of the atomic transitions, are defined as 

\begin{subequations}
	\label{detunings}
	\begin{align}
		\delta_{cw} &= \omega_{21} - \omega_{cw} - k_{cw}v,
		\\
		\delta_{m}  &= \omega_{32} - \omega_{m} + k_{m}v.
	\end{align}
\end{subequations}

\noindent where $k_{cw,m}$ are the wavenumbers of the cw laser and the m-th mode of the frequency comb.

We assume that the blue fluorescence is proportional to the population of the state $\left|3\right\rangle$, $\rho_{33}$. 
As the diode beam may be intense, we cannot apply second-order time-dependent perturbation theory, as in Ref.~\cite{Stalnaker2010}. The Bloch equations (\ref{bloch}) are solved exactly in the steady-state regime, with the help of a computer algebra system.

We plot in figure 6 the population $\rho_{33}$ as a function of (a) the Rabi frequency of the diode laser and (b) the interaction time, for systems with the $\left|1\right\rangle \rightarrow \left|2\right\rangle$ transition open (lower/blue curves)  or closed (upper/red curves). In Fig. 6(a), it is clear the difference in the population for high diode laser intensities, when the optical pumping becomes important. We also noticed a small difference in the values of the saturation Rabi frequency for the two systems. Figure 6(b) is another way of visualizing the distinction between them, since the $\rho_{33}$ values are different near the region of $\Gamma/2\pi=200$ kHz, which corresponds to the experimental conditions. Other parameters used: $\rho_{11}^0 = 1$, $\gamma_{22}/2\pi = 6$ MHz, $\gamma_{33}/2\pi = 0.66$ MHz, $\gamma_{12} = 0.5\gamma_{22}$, $\gamma_{13} = 0.5\gamma_{33}$, $\gamma_{23} = 0.5(\gamma_{33}+\gamma_{22})$ and $\Omega_m = \gamma_{33}$. 

\begin{figure}[ht]
    \centering
  \includegraphics[width=8.5cm]{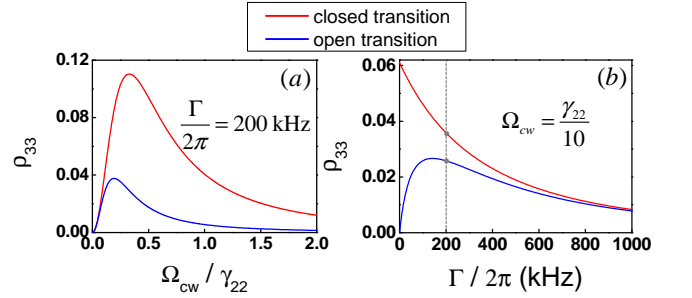}
  \caption {(Color online) Population $\rho_{33}$ as a function of (a) the diode Rabi frequency ($\Omega_{cw}$) and (b) the interaction time ($\Gamma$) for open (lower/blue curves) and closed (upper/red curves) systems in the  $\left|1\right\rangle \rightarrow \left|2\right\rangle$ transition driven by the diode laser.  The two fields are in resonance with their respective transitions, and for both curves $\Omega_{m}=\gamma_{33}$. }
  \end{figure} 

\subsection{\label{modeling}Modeling and velocity selectivity}

To compare with the experimental spectra we need to add the contributions due to all allowed two-photon transitions ($F \rightarrow F' \rightarrow F''$):

\begin{equation}
\label{rho33-calc}
\rho^{calc}_{33}(\delta) = \sum_{F,F',F''} \rho^{(F,F',F'')}_{33}(\delta).
\end{equation}

The matrix element of the electric dipole operator of each transition $F \rightarrow F'$ ($ \left\langle F \left| e\hat{\textbf{r}}\right| F' \right\rangle $) is calculated using an average over all matrix elements involving the magnetic sublevels $m_{F}$ and $m_{F'}$, which are obtained by the Wigner-Eckart theorem and the Clebsch-Gordan relations \cite{angularmomentum1960}:

\begin{eqnarray}
	\label{dipole-moments}
	\hspace{-20mm}&&\left\langle F,m_F \left| e\hat{r}_q \right| F',m_{F'} \right\rangle = \nonumber\\
	&=&\hspace{-1mm} (-1)^{2F' + I + J + J' + L + S + m_F + 1} \left\langle L \left\| e\hat{\textbf{r}} \right\| L' \right\rangle \nonumber\\
	&\times&\hspace{-1mm} \sqrt{(2F+1) (2F'+1) (2J+1) (2J'+1) (2L+1)} \nonumber\\
	&\times&\hspace{-1mm} \left( \begin{array}{ccc}F'&1&F\\m_{F'}&-q&-m_{F} \end{array} \right)\hspace{-1mm} \left\{ \begin{array}{ccc}J&J'&1\\F'&F&I \end{array} \right\}\hspace{-1mm} \left\{ \begin{array}{ccc}L&L'&1\\J'&J&S \end{array} \right\}. \nonumber\\
\end{eqnarray}
%

 
\noindent $\left\langle L \left\| e\hat{\textbf{r}} \right\| L' \right\rangle$ is the reduced matrix element that do not depend on the magnetic sublevels or the total angular momentum ($F$, $F'$, or $F''$) and $\hat{r}_q$ is the $q$ component of the vector operator $\hat{\textbf{r}}$ in the spherical basis. The $\left( \right)$ and $\left\{ \right\}$ terms are the  \textit{3-j} and \textit{6-j} Wigner symbols, respectively. 

For a specific transition $F \rightarrow F' \rightarrow F''$, we calculate 
the population $\rho_{33}$ of the final $F''$ state, in the steady-state regime, by solving the Bloch
equations for one specific diode frequency ($\delta=\delta_{cw}$) integrated over 
the contribution of all velocity groups ($\Delta=k_{cw}v$) within the Doppler profile:

\begin{equation}
\label{maxwell-boltzmann}
\begin{split}
\rho^{(F,F',F'')}_{33}(&\delta) = \frac{1}{\left( 0.36\pi\Delta^2_D \right)^{1/2}}\\
\times \int^{\infty}_{-\infty} &\rho_{33}(\Delta,\delta;\; \Omega_{F,F'}, \Omega_{F',F''})e^{- \Delta^2/0.36\Delta^2_D} d \Delta,\\
\end{split}
\end{equation}

\noindent with $\Delta_D$ being the inhomogeneous Doppler linewidth. In the calculations, the strength of a specific $F_i \rightarrow F_j$  transition is parametrized by the corresponding Rabi frequency: $\Omega_{F_i,F_j} = ({\mu_{F_i,F_j}{\cal E}})/{\hbar}$,
where $\mu_{F_i,F_j}$ is the electric-dipole moment for the transition and in the direction 
of the electric field, whose amplitude ${\cal E}$ may refer to the diode laser $({\cal E}_{cw})$ or a single
mode of the frequency comb $({\cal E}_{m})$, depending on the transition.
As we do not know the carrier-envelope-offset frequency, 
the frequency of the mode that drives the $F' \rightarrow F''$ transition 
is determine by the two-photon resonance condition combined with the diode frequency and the velocity of the atoms
at resonance. For an atom with velocity $v$, the frequencies of the diode laser ($5S \rightarrow 5P$ transition) and of the mode $m$ of the frequency comb ($5P \rightarrow 5D$ transition) are different, which gives a two-photon linewidth of approximately $\gamma_{33}+\gamma_{22}(\omega_{m}+\omega_{cw})/\omega_{21}$\cite {Liao1976}.

We assume that the two laser beams are counterpropagating, with perpendicular polarizations, as in the experiment, and calculate the population of the final state $\left|3\right\rangle$, $\rho^{(F,F',F'')}_{33}$, for each of the eight allowed two-photon transitions of the $^{87}$Rb, starting from the Doppler line $F=2$. We consider also that the cw laser is absorbed according to the Beer law, so that its intensity varies within the Doppler profile:

\begin{equation}
	\label{beer-law}
	{\cal E}_{cw} = {\cal E}^0_{cw}\text{exp}\left( -\alpha z e^{- \delta^2/0.36\Delta^2_D} \right),
\end{equation}

\noindent where $\alpha z$ is the optical density and ${\cal E}^0_{cw}$ simulates the field amplitude at the entrance of the Rb cell.

The results are shown in Fig. 7. We consider that the group of atoms with $v = 0$ is simultaneously in resonance with the diode laser at $F=2 \rightarrow F'=3$ transition and with the mode $m$ at the $F'=3 \rightarrow F''=4$ transition. The integral of Eq. (\ref{maxwell-boltzmann}) was calculated numerically by considering $\Delta_D/2\pi = 590$ MHz and the values of the field amplitudes, obtained from the experimental conditions, were: ${\cal E}_{cw}^0=350$ V/m  ($\Omega_{cw}\approx 2.7\gamma_{22}$ at $F=2 \rightarrow F'=3$) and ${\cal E}_{m}=350$ V/m ($\Omega_{m}\approx 6.6\gamma_{33}$ at $F'=3 \rightarrow F''=4$). The diode beam absorption was considered by assuming $\alpha z = 1.5$. As the two absorbed photons do not have the same frequency, different group of atoms are excited depending of the intermediate pathway. An example is the two peaks, almost superimposed, labeled by $2 \rightarrow 3$ in figure 7 (blue and green curves), corresponding to the transitions $F=2 \rightarrow F'=3\rightarrow F''=3$ and $F=2 \rightarrow F'=2\rightarrow F''=3$. The frequency difference for these two transitions is  of order of $\delta _{F'}=1.4$ MHz, and is given by the general expression \cite {Liao1976}: 

\begin{equation}
\label{difference-frequency}
\delta _{F'}=\triangle _{F'}\left( \frac{\lambda_{cw}}{\lambda_{m}}-1 \right),
\end{equation}

\noindent where $\triangle_{F'}$ is the frequency difference between the two hyperfine levels, $F'$, that participate of each pathway. 

\begin{figure}[ht]
  \centering
  \includegraphics[width=7.5cm]{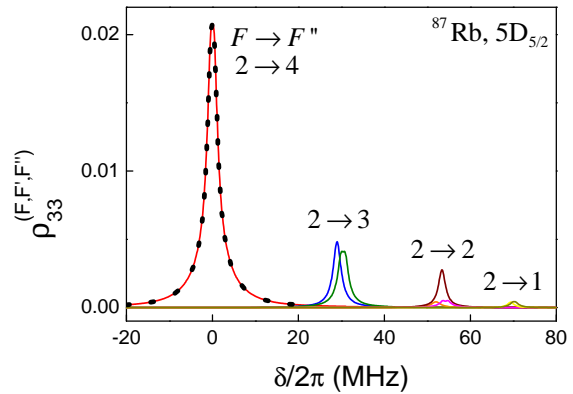}
  \caption {(a) (Color online) Population $\rho^{(F,F',F'')}_{33}$ as a function of the diode laser detuning for all allowed two-photon transitions of the $^{87}$Rb, starting from the Doppler line $F=2$. We consider that the group of atoms with $v=0$ is simultaneously in resonance with the diode laser at the $F=2 \rightarrow F'=3$ transition and with the mode $m$ at the $F'=3 \rightarrow F''=4$ transition. The dotted curve represents $\rho_{33}$ for the $F=2 \rightarrow F'=3 \rightarrow F'' = 4$ transitions calculated via fourth-order Runge Kutta method.}
  \end{figure} 

In fact, the greatest contribution for the upper level population, $\rho_{33}$, comes from the closed transition,  $F = 2 \rightarrow F' = 3 \rightarrow F'' = 4$. We use this transition to analyze the response of the system obtained in the time domain, considering the train of pulses given by Eq. (\ref{fs-field}). In this case, the Bloch equations are numerically solved in time by the classical fourth-order Runge-Kutta method, for each diode frequency and for each atomic group velocity. As this computation is very time consuming, we calculated the density matrix elements for all atomic group velocities in parallel using multiple threads of a graphic processing unit (GPU) \cite {Dem2013}. In this case, we have considered $N = 1000$ pulses, $T_R = 1$ ns, $T_p = 100$ fs (temporal linewidth of the pulses) and ${\cal E}_{fs}(0) = 3.5\times 10^{6}$ V/m. The response from the time domain calculations is represented by the black dots in Fig. 7, which appears over the curve labeled by $2 \rightarrow 4$ (red curve) obtained from the frequency domain calculations. The agreement between the two results is excellent and indicates how good is the approximation of working with a single mode of the frequency comb. This fact is due to the coherent accumulation process, in the upper transition, determined by constructive and destructive interferences between the electric field of the train of pulses and coherences excited by it. In particular, under the high repetition rate of the fs laser, 1 GHz, the necessary condition for the coherent accumulation of population and coherence is much better fulfilled.
  
  Another interesting feature is the strong dependence of the intensity relation between the peaks showed in Fig. 7 ($F=2 \rightarrow F''$) on the atomic-velocity group that participates on the two-photon transition. This dependence can be seen in Fig. 8 where we compare theory and experiment for the fluorescence signal of the $^{87}$Rb, $F=2$. We plot, in the same picture [figure 8(a)], several curves corresponding to the fluorescence signal obtained for different values of the repetition rate, as a function of the diode laser detuning. For a better comparison we subtracted the signal due only to the fs laser (background). The upper curve is the saturated absorption signal of the diode laser and is the same for all fluorescence signals of the picture. The first fluorescence curve at left (the black set of peaks) was obtained for $f_{R}=f^{0}_{R}=1.004\:396\:000$ GHz ($\triangle f_{R}=0$). In this case, we can see four peaks corresponding to the four possible hyperfine levels ($F''$) that can be excited starting from $F=2$, and all with almost the same intensity. As we vary the repetition rate ($\triangle f_{R}=300$ Hz) other atomic-velocity groups will be at resonance, so the corresponding fluorescence curve (the magenta set of peaks) will appear at a different diode frequency, dislocated in the saturated absorption curve.  The variation of $\triangle f_{R}$ between the curves is of 300 Hz and the total variation ($\triangle f_{R}=2700$ Hz) corresponds to change the optical diode frequency of almost 1 GHz. 
  
  The hyperfine levels of the intermediate state $5P_{3/2}$ have a strong influence in this intensity relation between the peaks. As we saw in Fig. 7, depending of the final level $F''$, different hyperfine levels $F'$ can be contributing for the same two-photon peak. For example, the resonant two-photon transition $F=2 \rightarrow F''=4$ involves only one pathway ($F=2 \rightarrow F'=3$), whereas the transition $F=2 \rightarrow F''=3$ has the contribution of two pathways ($F=2 \rightarrow F'=2,3$). In addition to the influence of the value of the electric dipole moment, we know that the more atoms are being excited the greater the peak amplitude of the two-photon transition. This behavior is in agreement with the results showed in Fig. 8(a), since at the left side of the Doppler curve, the transitions $F=2 \rightarrow F''=1,2$, with smaller electric dipole moment, are in resonance with more group of atoms (near the frequency transition $F=2 \rightarrow F'=1,2$ with the diode laser), making possible the observation of the four peaks. However, at the right side of the Doppler curve these transitions are far apart and so these peaks do not appear.
  
\begin{figure}[ht]
  \centering
  \includegraphics[width=8.5cm]{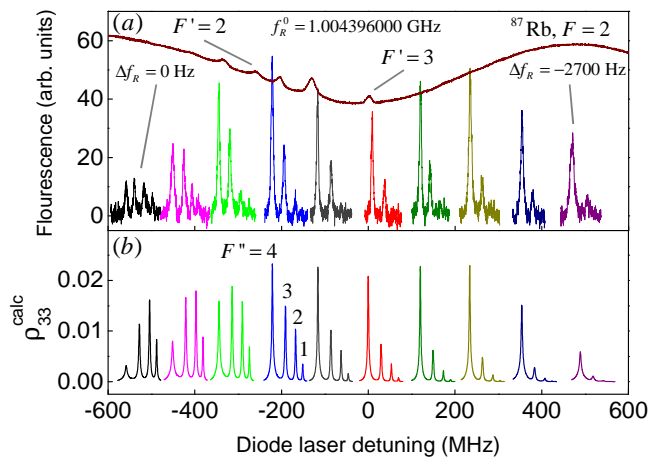}
  \caption {(Color online) (a) Curves of the fluorescence signal, for different values of repetition rate of the femtosecond laser, as the diode laser scan the $^{87}$Rb $F=2$ line (each color corresponds to one value of the repetition rate). The upper curve is the saturated absorption signal of the diode laser. (b) Excited state population, $\rho_{33}$, calculated from Eq. (\ref{rho33-calc}), for the same values of the repetition rate in
   (a).}
  \end{figure} 

In figure 8(b) we plot the excited state population, $\rho_{33}$, calculated from Eq. (\ref{rho33-calc}), as a function of the diode laser detuning, for the same values of the repetition rate presented in figure 8(a). In this modeling the values of the electric field of the two lasers were calculated from the power measured during the experiment. As we can see, our theoretical model not only well describes the intensity relation between the peaks for the same atomic-velocity group but also gives a good intensity relation between peaks of different atomic-velocity groups. The description is better near the center of the Doppler line where the contribution for the excitation process is predominantly due to atoms with $v=0$. The discrepancy is mainly on the sides of the Doppler line, where the experimental peaks present a small broadening. The average linewidth of the peaks corresponding to the $F=2 \rightarrow F''=3,4$ transitions as a function of their position within the Doppler profile is displayed in Fig. 9. At the center, the diode power is of order of $\mu$W (${\cal E}_{cw}$ $\approx$ 100 V/m), due to the strong absorption of the diode beam near the resonance, giving a linewidth around 5 MHz, whereas on the sides of the Doppler profile the linewidth tends to almost 10 MHz, indicating  a power broadening contribution. Other effects that also contribute to the broadening are: (i) the linewidth of the diode laser ($\sim$ 1 MHz), (ii) the jitter of the off set frequency in one scan, (iii) the Zeeman sublevels, and (iv) the broadening due to the time response of the photomultiplier. All these effects were not \textbf{taken }into account in our theoretical model.

\begin{figure}[ht]
  \centering
  \includegraphics[width=8.0cm]{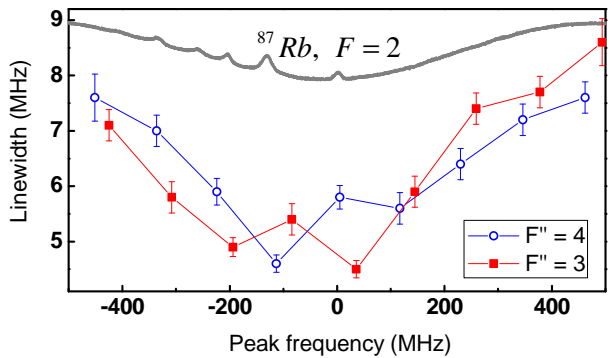}
  \caption {(Color online) The average linewidth of the peaks corresponding to the $F=2 \rightarrow F''=3$ (squares) and $F=2 \rightarrow F''=4$ (open circles) transitions as a function of their position within the Doppler profile. The upper curve is the saturated absorption curve of the diode laser.
 }
  \end{figure} 



\subsection{\label{OP} Influence of the optical pumping}

An interesting situation occurs when one group of atoms is in resonance, simultaneously, with an one-photon transition, driven by the diode laser, and with a two-photon transition, driven only by the femtosecond laser. In this case, the optical pumping induced by the diode laser can contribute to an increase or decrease of the background signal of the blue fluorescence which is only due to the two-photon transition by the fs mode-locked laser. An example of this effect is shown in Fig. 10,  where the bottom curve represents the fluorescence signal as a function of the diode laser frequency detuning for an average of ten scans processed by the oscilloscope. We clearly see two broad peaks, with linewitdhs of order of 29 and 44 MHz, respectively, separated by $\sim 157$ MHz, which are due to the optical pumping by the diode laser. For these conditions, one of the peaks corresponds to the fluorescence of a group of atoms that is in resonance, simultaneously, with the diode laser at the one-photon transition $F=2 \rightarrow F'=2$ and with the fs laser at $F=1 \rightarrow F''$ two-photon transition; whereas, the other large peak comes from the other group of atoms that is in resonance with the diode laser at the $F=2 \rightarrow F'=1$ transition (separated by 157 MHz from the other transition).

\begin{figure}[ht]
  \centering
  \includegraphics[width=8cm]{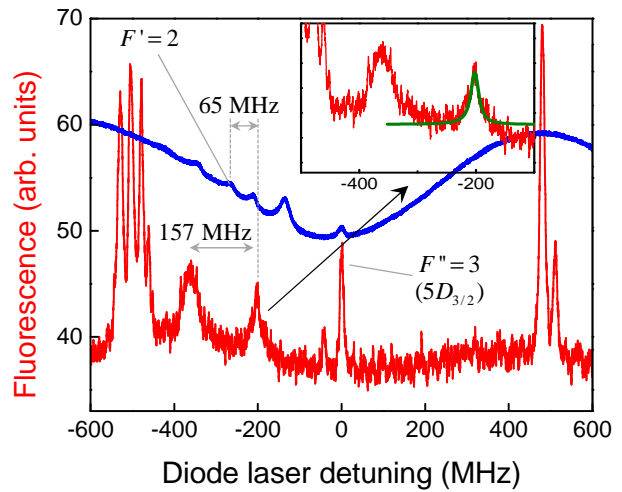}
  \caption {(a) (Color online) Fluorescence at 420 nm as a function of the diode laser detuning for $^{87}$Rb, $F=2$ (bottom/red curve). The narrow peaks in the center and the sides correspond to the transitions $5S_{1/2} \rightarrow 5D_{3/2}$ and $5S_{1/2} \rightarrow 5D_{5/2}$, respectively. The broad peaks indicated by a separation of 157 MHz are that induced by the diode optical pumping. The curve is an average of ten scans. The upper (blue) curve is the saturated absorption. For this scan $f_{R}=1.004\:382\:780$ GHz and $f_{0}=692$ MHz (see text). The solid/green line in the inset is the calculated $\rho_{33}$ population, using ${\cal E}_{cw}^0=500$ V/m, ${\cal E}_{m}=350$ V/m and ${\cal E}_{m'}=400$ V/m. }
  \end{figure} 

The Bloch equations, Eqs. (\ref{bloch}), can be also used to model the shape of the peaks induced  by optical pumping. For this, we need to take into account two modes, $m$ and $m'$, of the frequency comb, in a co-propagating configuration, and consider that the cw field of the diode laser contributes only to modify the population of state $\left|1\right\rangle$. A schematic representation of the energy levels is shown in Fig. 11. The $\left|a\right\rangle$ and $\left|b\right\rangle$ states are not included in the Bloch equations, and they are only used to describe the ground state population changes due to the optical pumping. So that, we solve Eqs. (\ref{bloch}) replacing the field of cw laser by the field of $m'$-th mode and the ground state population

\begin{equation}
\label{rho11-modification}
\rho^{(0)}_{11}\rightarrow\rho^{(0)}_{11}+\left[1+\frac{\gamma^{2}_{bb}+4(\Omega^{2}_{cw}+\delta^{2}_{cw})}{2\Omega^{2}_{cw}(\gamma_{bb}/\Gamma)}\right]^{-1}.
\end{equation}

The second term of the right side of Eq. (\ref{rho11-modification}) is the solution of the Bloch equations in the steady-state regime for $\rho_{11}$ in a $\Lambda$-type three-level system ($\left|a\right\rangle$, $\left|b\right\rangle$ and $\left|1\right\rangle$, Fig. 11) when the cw laser is close to the $\left|a\right\rangle \rightarrow \left|b\right\rangle$ transition in the approximation $\Gamma \ll \Omega_{cw},\gamma$ \cite{Polo2011}. The position of these large peaks into the Doppler profile also influences their shapes and can be determined from $f_{R}$, $f_{0}$, and the detunings of the one and two-photon resonances. 

\begin{figure}[ht]
  \centering
  \includegraphics[width=5.0cm]{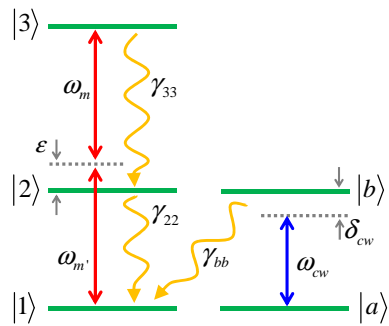}
  \caption {(a) (Color online) Theoretical model to describe the peaks induced by the diode optical pump. States $\left|a\right\rangle$ and $\left|b\right\rangle$ (not included in the Bloch equations) are used in Eq. (\ref{rho11-modification}) to modify the population of state $\left|1\right\rangle$. $\epsilon$ represents the detuning of the modes $m$ and $m'$ in relation to the transition with the intermediate level.} 
  \end{figure}

Although we have not measured the offset frequency of the femtosecond laser we can estimate its value, for a set of scans, by using the frequency values of the atomic transitions that are very well known. For that, we select a peak in the blue fluorescence signal related to a specific two-photon transition and use the fact that this signal is generated by a specific group of atoms that satisfy the resonant conditions, for one- and two-photons, given by Eqs. (\ref{detunings}) with $\delta_{cw}=\delta_{m}=0$. For the experimental curve of Fig. 10, the offset frequency is calculated from the central peak, where a group of atoms with $v = 0$ is at resonance with the diode laser in the $F = 2 \rightarrow F' = 3$ transition and with the $m$-th mode of the frequency comb in the $F' = 3 \rightarrow F'' = 3$ ($5D_{3/2}$) transition ($\omega_{32}$) and, from the repetition rate of the fs laser, $f_{R}$=1.004\:382\:780 GHz:

\begin{equation}
	\label{off-set}
	f_0 = \dfrac{\omega_{32}}{2\pi} - \left\lfloor \dfrac{\omega_{32}}{2\pi f_R} \right\rfloor f_R,
\end{equation}

\noindent given a value of $f_{0}=692$ MHz. The symbol $\left\lfloor \:\: \right\rfloor$ represents the integer part of a number.

In the inset of Fig. 10, the solid line (green curve) over the experimental curve represents the population $\rho_{33}$ calculated, as in Fig. 7, with the modifications described above. From this procedure we obtain the background signal, determined by the two-photon transition driven by the modes of the frequency comb due to the interaction with the groups of atoms that are far off resonance with the diode laser. We conclude, from the position of the broad peak with respect to the Doppler profile and from the estimated value of the $f_0$, that the main contributions comes from the transitions $^{87}$Rb, $F = 1 \rightarrow F' = 0 \rightarrow F'' =1 $ ($5D_{5/2}$) with the modes $m$ and $m'$ and $F = 2 \rightarrow F' = 2$ with the cw field; having the detuning of the one-photon transition of $\epsilon/2\pi = -35$ MHz for the atomic-velocity group at 65 MHz.

In the approximation of $\Gamma_{cw}\ll\Omega_{cw}, \gamma_{bb}$ , Eq. (\ref{rho11-modification}) gives a linewidth of 

\begin{equation}
\label{linewidth-pump-peaks}
\triangle\omega=\Omega_{cw}(2\gamma_{bb}/\Gamma_{cw})^{1/2}.
\end{equation}

\noindent From Eq. (\ref{linewidth-pump-peaks}), we get $\triangle\omega/2\pi = 15$ MHz, less than the experimental linewidth (29 MHz). As $f_{R}$ is fixed, the main mechanism that increases the peak linewidth are the fluctuations during the average of ten scans, caused mainly by the free carrier-envelope-offset frequency, $f_{0}$.


\section{\label{sec:4}Conclusion}
  
  In this paper we have presented new results on the velocity-selective two-photon absorption induced by the combined action of a diode laser and a train of ultrashort pulses. The blue fluorescence was investigated when the cw-laser frequency, at a fixed repetition rate of the fs laser, is scanned over the $D_{2}$ Doppler lines of $^{85}$Rb and $^{87}$Rb, showing a large spectral range that includes the $5D_{3/2}$ and $5D_{5/2}$ states. The  resolution is limited by the free-running offset frequency and the linewidth of the diode laser, without the need of direct spectral filtering of the fs laser.  By varying the value of the repetition rate we were able to study the response of different atomic-velocity groups. Moreover, the large frequency separation of 1 GHz of the modes of the frequency comb, allowed us to distinguish the hyperfine levels of the $5D$ excited state and their relative intensities for different atomic groups within the Doppler profile. A good quantitative description of the experimental spectra at different positions of the Doppler line is provided in the frequency domain picture, taking into account the high intensity and the absorption of the diode laser, as well as, the Maxwell-Boltzmann population distribution. We have also shown that the results of the  numerical solution of the Bloch equations for a three-level system interacting with both a cw laser and the whole train of ultrashort pulses, fits well our experimental data, and confirm our analytical treatment based in a single mode of the frequency comb.  Further, our results also revealed the contribution of the optical pumping to the two-photon transition driven only by the fs laser, which is well described by considering the interaction of the atomic system with two modes of the frequency comb.
 
\begin{acknowledgments}
This work was supported by CNPq, CAPES, FACEPE and FAPERO (Brazilian Agencies).
\end{acknowledgments}



\end{document}